\documentclass[twocolumn, prb, superscriptaddress]{revtex4-1}
\usepackage{amsmath}
\usepackage{color}
\usepackage{amssymb}
\usepackage{graphicx}
\usepackage{epstopdf}
\usepackage{verbatim}
\pdfoutput=1
\newcommand{\be}{\begin{equation}}
\newcommand{\ee}{\end{equation}}

\definecolor{darkblue}{rgb}{0.0,0.0,0.55}

\begin{document}

\title{Correlation effects in the capacitance of a gated carbon nanotube}

\date{\today}
\author{Han Fu}
\author{B. I. Shklovskii}
\affiliation{Fine Theoretical Physics Institute, University of Minnesota, Minneapolis, Minnesota 55455}

\author{Brian Skinner}
\affiliation{Materials Science Division, Argonne National Laboratory, Argonne, IL  60439, USA}

\begin{abstract}

For a capacitor made of a semiconducting carbon nanotube (CNT) suspended above a metallic gate, Coulomb correlations between individual electrons can lead to a capacitance that is much larger than the geometric capacitance. We argue that when the average spacing $n^{-1}$ between electrons within the low density 1-dimensional electron gas (1DEG) in the CNT is larger than the physical separation $d$ between the CNT and the gate, the enhancement of capacitance is expected to be big.  A recent experiment,\cite{waissman_realization_2013} however, has observed no obvious increase of capacitance even at very low electron density. We show that this smaller capacitance can be understood as the result of the confining potential produced by the potential difference between the source/drain electrodes and the gate, which compresses the 1DEG when the electron number decreases. We suggest that by profiling the potential with the help of multiple split gates, one can return to the case of a uniform 1DEG with anomalously large capacitance.

\end{abstract} \maketitle

\section{Introduction}

One of the great features of low-dimensional materials is that their properties can be continuously altered using electrostatic gating.  Such gating allows one to tune the bulk electron density widely and reversibly, and thereby probe different regimes of electronic behavior simply by turning an experimental knob.  What's more, the material's response to the applied voltage gives information about the electronic compressibility, which is in turn a reflection of the nature of the interactions and correlations between electrons.  For example, for a gated two-dimensional electron gas (2DEG), the two-dimensional electron density $n_2$ can be tuned by changing the gate voltage $V$, and the resulting differential capacitance $C \propto dn_2/dV$ of the gate-2DEG system can be used as a probe of quantum and many-body correlated electron physics.  The general promise of capacitance as a probe of low-dimensional materials has been appreciated for several decades \cite{bello_density_1981, luryi_quantum_1988, kravchenko_evidence_1990, eisenstein_compressibility_1994, eisenstein_negative_1992, shapira_thermodynamics_1996}, and has recently been used as the basis for a number of dramatic observations in graphene \cite{hunt_massive_2013, yu_interaction_2013, feldman_unconventional_2012}.

One particularly striking observation is that electron correlations can lead to an enhancement of the capacitance above the normal ``geometric" value $C_g$.  For example, for a low-density 2DEG, the capacitance has been predicted to follow the relation \cite{bello_density_1981}
\be
(C/A)^{-1} = (C_g/A)^{-1} - 1.5 n_2^{-1/2}
\label{eq:C}
\ee
(in Gaussian units), where $A$ is the capacitor area.
The second term on the right hand side of Eq.\ (\ref{eq:C}) arises as a consequence of spatial correlations between electrons, which lower the total energy of the electron gas relative to the uniform state and thereby drive up the capacitance.  While a similar result is often explained in terms of the exchange interaction and referred to as a ``quantum capacitance", its origin can be understood at the level of a classical electron crystal (a Wigner crystal), where positional correlations between electrons reduce the total electrostatic energy \cite{bello_density_1981, skinner_anomalously_2010}.  An analogous result to Eq.\ (\ref{eq:C}) has been observed already for a number of two-dimensional systems, including GaAs \cite{eisenstein_compressibility_1994, eisenstein_negative_1992}, the LaAlO$_3$/SrTiO$_3$ interface \cite{li_very_2011}, and graphene in a magnetic field \cite{skinner_giant_2013, skinner_effect_2013}.

Recently, however, it was pointed out that Eq.\ (\ref{eq:C}) cannot retain its validity at arbitrarily small electron density \cite{skinner_anomalously_2010}.  Indeed, at sufficiently small density that the typical distance between electrons $n_2^{-1/2}$ becomes much longer than the distance $d$ to the metal gate, Eq.\ (\ref{eq:C}) gives a negative result for $C$.  Such negative values of the capacitance are forbidden by thermodynamic stability criteria \cite{lifshitz_electrodynamics_1984} (assuming that the capacitance is measured in the low frequency limit).  This failure of Eq.\ (\ref{eq:C}) points to the appearance of new physics in the very low-density regime.  This physics was outlined for two-dimensional systems in Ref.\ \onlinecite{skinner_anomalously_2010}, where it was shown that at $n_2 d^2 \ll 1$ the interactions between electrons become strongly screened by image charges in the metal gate, so that the electron-electron repulsion takes the form of a dipole-dipole interaction. Such screening allows for an enormous enhancement of the capacitance over the geometric value in the low-density limit.  It also implies a re-entrance of the Fermi liquid phase at small $n_2$, and an ultimate saturation of the capacitance at $C/A \sim m e^2/\hbar^2$ in the limit $n_2 \rightarrow 0$, where $m$ is the effective electron mass \cite{skinner_simple_2010}, $e$ is the electron charge, and $\hbar$ is the reduced Planck constant.

So far, however, experiments have been largely unable to probe this low-density regime of ``dipolar electrons".  While some efforts have come close \cite{li_very_2011}, experiments on two-dimensional electron systems have been largely hindered by the ability to make very thin devices.  The presence of disorder is also a ubiquitous confounding factor, since disorder produces strong spatial modulation of the electron density when the electron number is low.

\begin{figure}[htb]
\begin{center}
\includegraphics[width=0.48\textwidth]{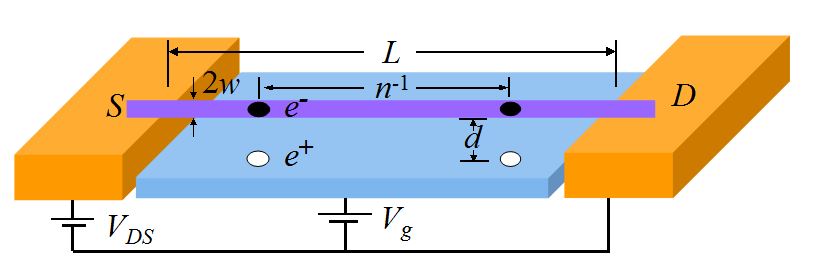}\\
\caption{(Color online) Experimental set-up of the CNT device. A CNT (purple) is suspended above a gate electrode (blue) between source and drain electrodes (orange). $L$ is the length of the CNT, $w$ is its radius, $d$ is the separation between the CNT and the metallic gate, and $n^{-1}$ is the average distance between electrons. $V_{DS}$ denotes the source-drain voltage, which is much smaller than the the gating voltage $V_g$. $\bullet$ symbols represent electrons and $\circ$'s denote the induced positive charges on the gate.  }\label{fig:1}
\end{center}
\end{figure}

Recently, however, developments in ultra-clean, gated carbon nanotube devices have shown the promise to overcome both of these difficulties\cite{waissman_realization_2013}.  In such one-dimensional devices, a single semiconducting CNT is suspended above a gate electrode, as illustrated in Fig.\ \ref{fig:1}, and its capacitance is measured at low temperature.  Importantly, existing devices exhibit a very large ratio of the length $L$ to the gate distance $d$, and can have their total electron number $N$ tuned incrementally from zero.  These conditions together raise the possibility of convincingly probing the dipole-interacting regime of electron behavior, in which the inter-electron separation is much larger than $d$.

The one-dimensional nature of a CNT, however, produces qualitative changes to the capacitance at low electron density that have so far not been studied theoretically.  In this paper we address this problem, describing the nature of electron correlations across all different regimes of the total electron number $N$ and deriving analytical expressions for the differential capacitance.  Our analytical results are confirmed with simple numeric calculations.  We focus everywhere on the case of a \emph{semiconducting} CNT, where electrons have a finite effective mass at low density.  The case of a metallic CNT, where electrons have a linear, Dirac-like dispersion relation, requires separate consideration.

In describing the behavior of the system, we focus on two cases for the experimental environment.  In the first case (Sec.\ \ref{sec:noconf}), we imagine that the system is spatially uniform, so that electrons are not subjected to any external potential, as in previous work \cite{skinner_anomalously_2010}. However, we find a big discrepancy between the predicted results for this case and the experimental data in Ref. \onlinecite{waissman_realization_2013}. A consideration of the experimental setup leads us to the second case (Sec.\ \ref{sec:withconf}), in which electrons are subjected to a confining electrostatic potential, as might arise naturally from the applied potential difference between source/drain and gate electrodes.  As shown in Sec.\ \ref{sec:withconf}, this confining potential is crucial for understanding previously-published experimental data.  In Sec.\ \ref{sec:variational}, we check our calculations using a variational method that properly accounts for the quantum kinetic energy ignored in Sec.\ \ref{sec:noconf} and \ref{sec:withconf}. The results are found to closely align with those of the previous sections, thereby justifying a simple classical description.  We conclude in Sec.\ \ref{sec:conclusion} by discussing the outlook for future experiments.  We also discuss how the case of a spatially-uniform system can be realized using a setup with multiple gates, which would allow for a significantly larger capacitance.

\section{Spatially-uniform case}
\label{sec:noconf}

In this section we consider a model in which the CNT is not subjected to any external potential, and its electron charge is simplified as a 1D uniform electron gas which forms a capacitor with the metallic gate placed below it. The CNT's radius $w$ is $\sim 1$ nm, much smaller than its length $L$ (which in Ref.\ \onlinecite{waissman_realization_2013} is $880$ nm) or the gate distance $d$ ($130$\,nm in Ref.\ \onlinecite{waissman_realization_2013}). The linear density of electrons along the nanotube is $n=N/L$, where $N$ is the total electron number. The typical distance between electrons in experiment is $n^{-1}=50-880$ nm. The relevant length scale for describing quantum mechanical effects is the effective Bohr radius, $a_B^* = \hbar^2/m e^2 \approx 8.9$ nm.
This large Bohr radius arises in experiment from stretching or curvature effects in the nominally metallic CNT, which open up a small band gap $\Delta=34$ meV, and then the effective mass of electrons is small: $m  \approx m_0/170$ (where $m_0$ is the bare electron mass).

Our general approach to calculating the capacitance (as outlined, for example, in Ref.\ \onlinecite{skinner_anomalously_2010}) is to first evaluate the total energy $E$ of the CNT-gate system and then calculate its second derivative with respect to the total electron number. In particular, the differential capacitance is given by the formula
\be
C=\left(\frac{d^2E}{e^2d N^2}\right)^{-1}=\left(\frac{d\mu}{e^2 d N}\right)^{-1} ,
\label{eq:Cdef}
\ee
where $\mu = dE/dN$ is the chemical potential, which includes interactions both between electrons within the CNT and between electrons and the positive charge in the gate.  Equation (\ref{eq:Cdef}) implies that capacitance can be thought of as equivalent to electron compressibility: large capacitance arises in situations where electrons are easily compressible.
We concentrate throughout this paper on the limit of zero temperature, which provides an accurate estimate for the capacitance whenever the thermal energy $k_BT$ is much smaller than the typical interaction energy per electron.  More generally, for finite temperature systems one can calculate the capacitance by replacing the energy $E$ in Eq.\ (\ref{eq:Cdef}) with the Helmholtz free energy.

\subsection{High Density Regime: $n\gg 1/a_B^*$}

For an interacting electron system, when the average distance between electrons is much smaller than the effective Bohr radius $a_B^*$, the electrons can be considered to be weakly interacting and are spread uniformly over the nanotube.  Thus, at sufficiently high density, the CNT can be modeled as a metallic wire, and the dominant contribution to the total energy $E$ comes from the electrostatic energy of the wire-gate system.  Such a system has total energy $e^2N^2\ln(2d/w)/L$, which corresponds to a chemical potential $\mu = dE/dN = 2e^2n\ln(2d/w)$,\cite{Jackson} so that the capacitance is
\begin{equation}
C\approx C_g=\frac{L}{2\ln(2d/w)}.
\label{eq:Cg}
\end{equation}
We refer to this quantity as the geometric capacitance.

One can note, however, that Eq. (\ref{eq:Cg}) gives an expression for the capacitance that is exact only when $n\gg1/\sqrt{a_B^*w}$.  At densities $1/a_B^* \ll n \ll 1/\sqrt{a_B^* w}$, though the electron gas is still in the weakly interacting regime, the argument inside the logarithm is modified due to short-range correlations between electrons.  Such correlations emerge at small enough electron separation $r$ that the Coulomb repulsion $e^2/r$ is larger in magnitude than the typical kinetic energy $\sim \hbar^2 n^2/m$, and they produce a truncation of the electron-electron interaction at short distances.  As a consequence, in this density range the argument of the logarithm is replaced by $d n^2 a_B^*$.  Since this replacement produces only a small correction, one can still say that Eq. (\ref{eq:Cg}) is correct to within a logarithmic factor whenever $na_B^* \gg 1$.

\subsection{Intermediate Regime: $1/a_B^* \gg n \gg 1/d$}

When $n \ll 1/{a_B^*}$, the typical Coulomb energy $\sim e^2 n$ dominates over the kinetic energy $\sim \hbar^2 n^2/m$, and the electrons assume a correlated state reminiscent of a Wigner crystal. (Indeed, such a Wigner-crystal-like state has already been reported experimentally in semiconducting CNTs \cite{Vikram_2008}.)  One can therefore estimate the total energy of the electron system by modeling it as a line of classical point charges with the gate-screened interaction $V(r) = e^2/r - e^2/\sqrt{r^2 + (2d)^2}$.  This approach gives
\be
E=n L\left\{-\frac{e^2}{4d}+\sum^\infty_{i=1}\left[\frac{e^2}{i/n}-\frac{e^2}{\sqrt{(i/n)^2+(2d)^2}}\right]\right\},
\label{eq:Eclassical}
\ee
where the first term inside the braces represents the attraction of a given electron to its own image charge, and the sum describes the repulsion between electrons.  This classical description, which ignores the quantum kinetic energy of electrons, is justified in detail in Sec.\ \ref{sec:variational}.

Evaluating Eq.\ (\ref{eq:Eclassical}) in the limit $n d^2 \gg 1$ and taking the derivative with respect to $n$ gives
\be
\mu \simeq e^2 \left[ 2n\ln(2dn)+0.768n-\frac{1}{4d} \right],
\label{eq:muint}
\ee
so that the capacitance
\begin{equation}
C\approx\frac{L}{2[\ln(2dn)+1.384]}.
\label{eq:Cint}
\end{equation}
This expression implies a marginal, logarithmic increase in the capacitance with decreasing electron density, $C/C_g \simeq 1 + [\ln(1/nw)-1.384]/[\ln(2 n d) + 1.384]$. Such a small correction to the geometric capacitance is similar to the situation in 2D, where electron correlations provide a small, positive correction to the geometric capacitance at intermediate electron density \cite{bello_density_1981, skinner_anomalously_2010}. This enhancement of the capacitance is a manifestation of electron correlations,  which allow the electrons to lower their energy for a given concentration by avoiding each other spatially, and thereby achieve a higher compressibility.

\subsection{Low Density Regime: $n \ll 1/d$}

In the regime of very low electron density, $n \ll 1/d$, the distance between neighboring electrons is much longer than the distance between an electron and its image charge in the metal gate. As a result, the electron-electron interaction takes the form of a dipole-dipole potential: $V(r) \simeq 2 e^2 d^2/r^3$.  Consequently, the total energy can be calculated as
\be
E \simeq n L\left[-\frac{e^2}{4d}+\frac{1}{2}\sum^\infty_{i=1}\frac{4e^2d^2}{(i/n)^3}\right].
\nonumber
\ee
Therefore, by Eq.\ (\ref{eq:Cdef}),
\be
\mu \simeq -\frac{e^2}{4d}+9.62e^2d^2n^3
\label{eq:mulow}
\ee
and
\begin{equation}
C \simeq \frac{0.035L}{d^2n^2}.
\label{eq:Csmall}
\end{equation}

Thus, when the electron density is made sufficiently low that $nd \ll 1$, the capacitance grows very quickly with decreased $n$, as in the 2D case \cite{skinner_anomalously_2010}. This large capacitance can be seen as a consequence of the screening of the electron-electron interactions by the metal gate, which renders them effectively short-ranged.  This truncation of the interaction, together with the positional correlations between electrons, allows the electrons to be highly compressible, and drives the capacitance far above the geometric value given by Eq.\ (\ref{eq:Cg}).

Finally, if the electron density is reduced so far that $n\ll a_B^*/d^2$, then the dipolar interaction between electrons becomes too weak to maintain a strong degree of spatial correlation between electrons.  In other words, one can think that the zero-point fluctuations of an electron in the confining potential created by its neighbors become comparable in amplitude to the spacing between electrons.  In this case, the Wigner-crystal-like correlations between electrons are melted, and the system can again be described as a weakly-interacting liquid.  In such a situation the total energy of the electron system is dominated by the kinetic energy, and the capacitance is proportional to $L/na_B^*$. This extreme low density regime is never reached in the experiments of Ref.\ \onlinecite{waissman_realization_2013}, and is neglected for the remainder of this paper in order to relate more closely to experiment.

\begin{figure}[h]
\begin{center}
\includegraphics[width=8.5cm, height=5.5cm]{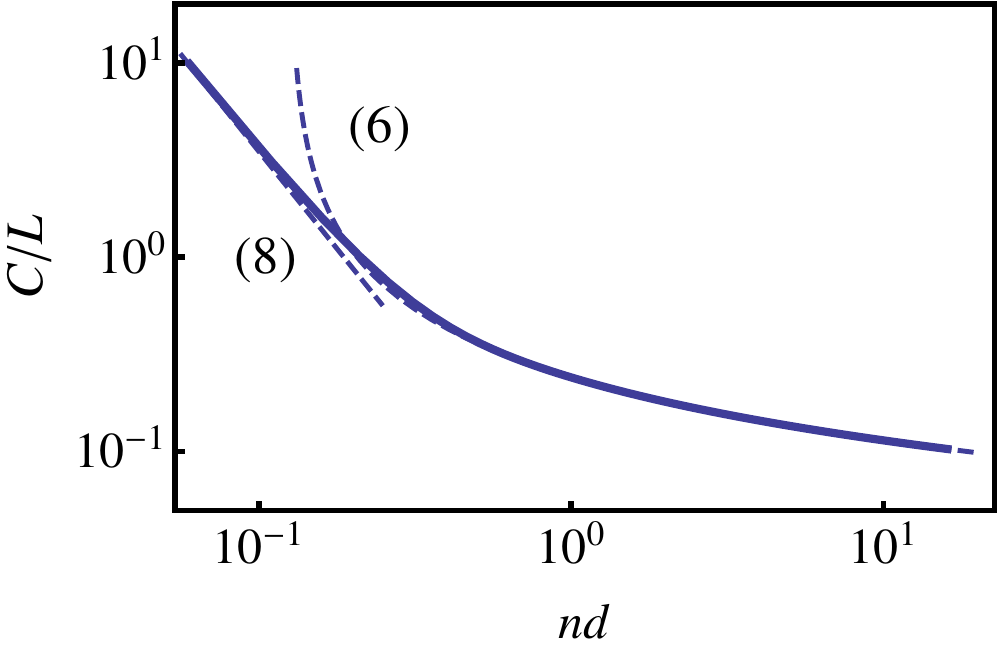}\\
\caption{Capacitance per unit length, $C/L$, for a CNT without a confining potential, plotted as a function of the electron density $n$ multiplied by the gate distance $d$. The solid line shows a numerical evaluation of $C$ using Eqs.\ (\ref{eq:Eclassical}) and (\ref{eq:Cdef}). Dashed lines illustrate the asymptotic behaviors of $C$ in different density limits. At small $n$, the capacitance is described by Eq.\ (\ref{eq:Csmall}). In the intermediate regime, the behavior is described by Eq.\ (\ref{eq:Cint}). At much larger $n$, $C$ saturates at the geometric value, Eq.\ (\ref{eq:Cg}).
}
\label{fig:2}
\end{center}
\end{figure}

The behavior of the capacitance across different electron density regimes is shown in Fig.\ \ref{fig:2}. The capacitance $C$ shows a very weak dependence on the electron density when $nd \gg 1$, while at $nd \ll 1$ the capacitance rises quickly with decreasing electron concentration.

Interestingly, the first experimental measurement of the capacitance of a single CNT capacitor observed a similar trend of increasing capacitance with decreasing electron density \cite{Ilani_2006}.  This experiment, however, reported only a small increase in the capacitance relative to the uncorrelated state, rather than the large effect depicted in Fig. \ref{fig:2}.  This absence of large capacitance enhancement in this experiment can likely be attributed to the relatively large experimental temperature ($T \approx 77$\,K), which destroys electron correlations when the electron density is low and the typical electron-electron interaction energy is weak.  In particular, for the experiment of Ref.\ \onlinecite{Ilani_2006}, the temperature was large enough that $k_BT \sim e^2/d$, which implies that electron correlations are lost at all $n \lesssim 1/d$ and the theory outlined in this section no longer holds.  Since the increase in capacitance is relatively weak at $n d \gtrsim 1$ (as shown in Fig.\ \ref{fig:2}), one can conclude that only a relatively small enhancing effect of the capacitance should be expected. Disorder effects may also contribute to the observed absence of large capacitance at low density.

Fortunately, more recent experiments\cite{waissman_realization_2013} have performed capacitance measurements at much lower temperature (4 K), for which strong electron correlations can persist down to low densities $n \ll 1/d$.  These experiments are considered in detail in the following section.

\section{Capacitance in the presence of a confining potential}
\label{sec:withconf}

In the previous section, we derived the dependence of the capacitance on the electron density for a CNT system without a confining potential, and we showed that $C$ rises sharply at small $n \ll 1/d$. A recent experiment\cite{waissman_realization_2013}, however, showed no such enhancement of the capacitance, even when $(N/L)d$ is made as small as $\approx 0.15$. In this experiment, the capacitance was studied by examining the Coulomb blockade in the source-drain current (rather than by measuring the impedance produced by a small AC modulation of the gate voltage, as is typical for 2D systems). The authors of Ref.\ \onlinecite{waissman_realization_2013} found that the spacing in the source-drain voltage $V_{DS}$ between successive Coulomb blockade conductance peaks was nearly constant, $\Delta V_g = 31.5 \pm 1.5$ mV, which suggests a constant capacitance $C_m = e/\Delta V_g \approx  46$\,nm.  This spacing was consistent over the reported range of electron number, $1 \leq N \leq 17$, which corresponds to an electron density $0.001\,\text{nm}^{-1} < n < 0.019\,\text{nm}^{-1}$.

\begin{figure}[h]
\begin{center}
\includegraphics[width=8.5cm, height=5.5cm]{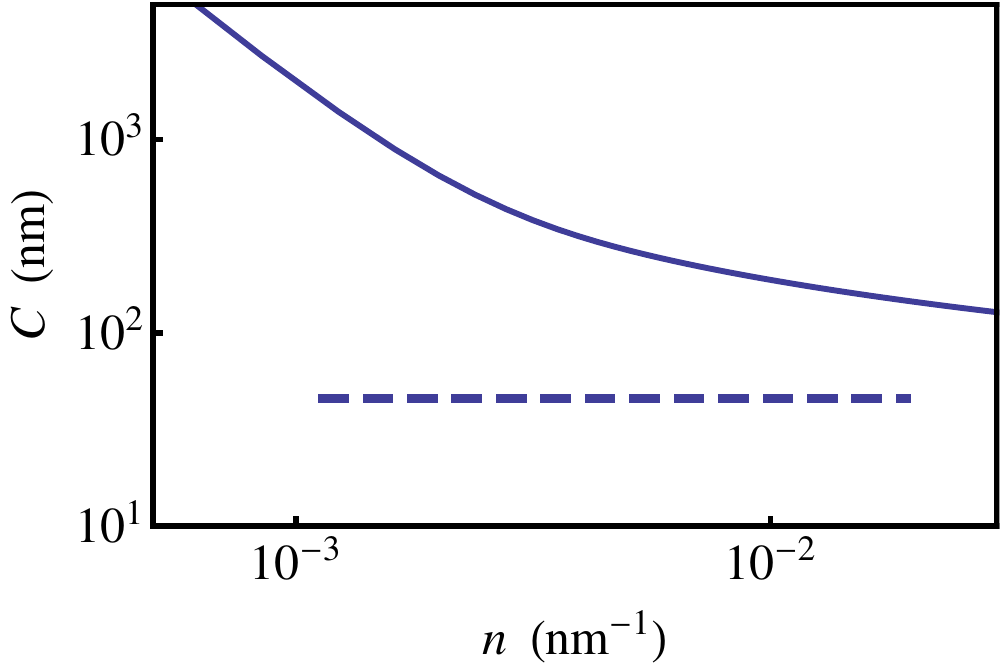}\\
\caption{Prediction for the capacitance $C$ as a function of the electron density $n$ using the experimental parameters of Ref.\ \onlinecite{waissman_realization_2013}. The solid line is the prediction of the theory for a uniform gas (see Sec.~\ref{sec:noconf}). The dashed line shows the actual capacitance $C_m = 45.7$\,nm measured experimentally.}
\label{fig:3}
\end{center}
\end{figure}

Inserting the experimental parameters from Ref.\ \onlinecite{waissman_realization_2013} into the theoretical prediction of the previous section gives the result shown in Fig.\ \ref{fig:3}.
As can be seen from this figure, this theory predicts a capacitance that is larger than the experimentally observed value by as much as $20$ times.  The experimental results also do not exhibit the sharp upturn in capacitance at small electron density that is characteristic of the theory in the previous section.

\begin{figure}[h]
\begin{center}$
\begin{array}{cc}
\includegraphics[width=0.48\textwidth]{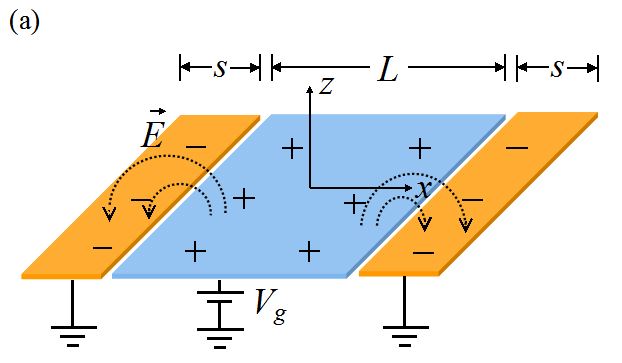}\\
\includegraphics[width=0.48\textwidth]{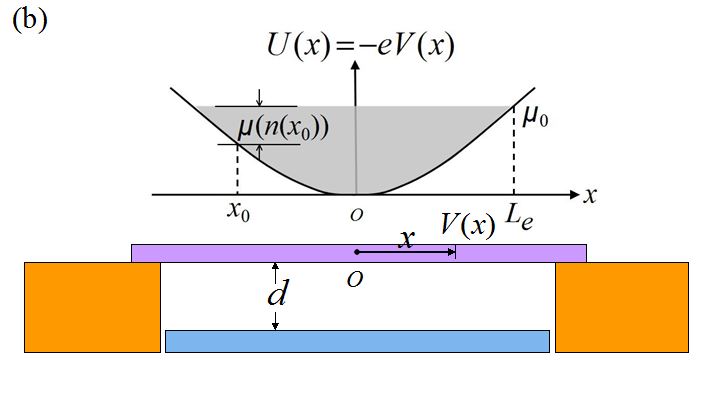}\\
\end{array}$
\caption{(Color online) Confining potential acting on electrons in the CNT formed by fields between source/drain electrodes and the gate.  (a) Schematic of the electric field produced by the potential difference between the source/drain electrode and the gate. The source-drain voltage $V_{DS}$ is neglected here. (b) This field produces a confining potential $U(x)$ along the CNT which leads to non-uniform electron density, so that the electrochemical potential $\mu_0=U(x)+\mu(n(x))$ is constant. The effective length $2L_e$ is defined by $n(\pm L_e)=0$.}
\label{fig:4}
\end{center}
\end{figure}

In order to resolve this discrepancy, we consider the effects of the confining potential produced by the electrodes.  In the experiment, a potential difference $V_g \approx 0.5-1$\,V is applied between the gate electrode and the CNT in order to bring the Fermi level close to the bottom of the conduction band.  This potential produces significant electric fields, as depicted in Fig.\ \ref{fig:4}, which imply that the electric potential along the CNT is non-uniform.

In order to estimate the strength of this confining potential, we solve for the electrostatic potential of a system of three coplanar metallic strips held at different voltages (due to the small size of the CNT, we can ignore its role in modifying the potential).  An analytic solution for the potential is presented in Appendix\ \ref{App:AppendixA}.  These metallic strips are assumed to have infinite length in the direction perpendicular to the CNT, but finite width, which in experiment is $s \approx 300$\,nm for the source/drain electrodes and $L = 880$\,nm for the gate.  This assumption of infinite length is justified by the large ratio of the electrodes' length to their width, which in Ref.\ \onlinecite{waissman_realization_2013} is $\ge 100$.  The assumption of coplanarity is justified by the small value of $d/L$.

If one defines the coordinate $x$ as the position along the CNT axis relative to the center of the CNT, then the electric potential $V$ has a parabolic maximum at $x = 0$, so that the confining potential can be written
\begin{displaymath}
U(x) = -eV(x,d) \simeq U(0)+\frac{e^2}{D^3}x^2
\end{displaymath}
where $U(0)$ is a constant and $D$ is a length scale that describes the strength of the confinement.  The expression for $D$ is given in Appendix\ \ref{App:AppendixA}, and for the experiments of Ref.\ \onlinecite{waissman_realization_2013}, $D \approx 72-90$\, nm.

When a small number of electrons are added to the CNT, the confining potential pushes these electrons to occupy a region near the center of the gate electrode, so that the electron-populated region has an effective length $2L_e$ that is shorter than the total CNT length $L$. As $N$ is reduced, electrons are more easily compressed by the confining potential and $L_e$ shrinks.  This reduction of the effective length of the capacitor tempers the increase of $C$ at low density that was derived in the previous section.

In the remainder of this section we calculate the capacitance in different regimes of the total electron number, taking into account the compression effect produced by the confining potential.  We focus our discussion here on the case where the local electron density is everywhere smaller than $1/a_B^*$ and larger than $a_B^*/d^2$, which is the relevant case for existing experiments.  Within this range of density, we focus on only two regimes:  $N/L_e\gg d^{-1} $ and $N/L_e\ll d^{-1}$, or equivalently (as shown below), $N\gg\left(D/d\right)^{3/2}$ and $N\ll\left(D/d\right)^{3/2}$.  These two regimes correspond, respectively, to the regime of a Wigner crystal-like state with unscreened interactions, and the regime of a ``dipole crystal'' with strongly-screened interactions.

\subsection{High Density Regime: $N\gg\left(D/d\right)^{3/2}$ }

Because of the confining potential, the electron density $n$ varies with position $x$ in order to maintain a constant value of the electrochemical potential $\mu_0=U(x)+\mu\left(n(x)\right)$ (see Fig.\ \ref{fig:4}).
In the regime of relatively high electron density, electron interactions are essentially unscreened by the gate, and the contribution to the chemical potential from electron interactions is given by Eq.\ (\ref{eq:muint}), $\mu\left(n(x)\right)\approx e^2n[2\ln(2dn)+0.768]$.  (Here, the $n$-independent term in $\mu$ is ignored, since this just redefines the reference point for $\mu$).
%
%
Thus, the electrochemical potential can be written
\be
\mu_0=\frac{e^2}{D^3}x^2+n(x)e^2[2\ln(2dn(x))+0.768].
\nonumber
\ee
Rearranging this expression for $n(x)$ gives
\be
n(x) = \frac{\mu_0/e^2 - x^2/D^3}{2\ln(2dn(x))+0.768} \, .
\nonumber
\ee
A solution can be obtained for $n(x)$ by taking successive approximations for the slowly-varying logarithm $\ln(2 d n(x))$; we treat this factor first as a constant of order unity and later replace $n(x)$ inside the logarithm with the calculated average value $N/(2L_e)$.
Since
\be
N=2\int^{L_e}_0 n(x) dx,
\nonumber
\ee
this approach gives
\begin{eqnarray}
L_e & \simeq & D\left\{N\ln\left[\frac{1.23N}{(D/d)^{3/2}}\right]\right\}^{1/3}, \nonumber \\
\mu_0 & \simeq &\frac{e^2}{D}\left\{N\ln\left[\frac{1.23N}{(D/d)^{3/2}}\right]\right\}^{2/3}.
\nonumber
\end{eqnarray}
Thus, as announced earlier, $N /L_e\gg d^{-1}$ is equivalent to $N\gg\left(D/d\right)^{3/2}$, and the resulting capacitance
\begin{equation}
C \simeq \frac{1.5}{\left\{\ln\left[\frac{1.23N}{(D/d)^{3/2}}\right]\right\}^{2/3}} d \left(\frac{D}{d}\right)^{3/2} \left[\frac{N}{(D/d)^{3/2}}\right]^{1/3}.
\label{eq:ChighD}
\end{equation}


One can notice that Eq.\ (\ref{eq:ChighD}) represents a qualitatively different dependence of the capacitance on electron number than the corresponding expression for the uniform gas, Eq.\ (\ref{eq:Cint}).  Rather than weakly increasing as the number of electrons in the CNT is reduced, Eq.\ (\ref{eq:ChighD}) implies that the capacitance \emph{decreases} with decreased $N$.  This decline of the capacitance is a result of the confining potential, which causes the effective capacitor length to shrink, and this shrinking of the effective length dominates over the weak effect of electron-electron correlations.

\subsection{Low Density Regime: $N\ll \left(D/d\right)^{3/2}$}

On the other hand, in the regime where electron-electron interactions are strongly screened by the metal gate, the effect of electron-electron correlations is large, and competes with the shrinking effective length to determine the overall capacitance.  In this regime, the contribution to the electrochemical potential arising from electron-electron interactions is given by Eq.\ (\ref{eq:mulow}), so that the electrochemical potential is
\be
\mu_0=\frac{e^2}{D^3}x^2+9.62e^2d^2n^3(x).
\nonumber
\ee
Solving this expression for the electron density gives
\be
n(x)=\left(\frac{\mu_0 - e^2 x^2/D^3}{9.62e^2d^2}\right)^{1/3}.
\ee

Using the expressions $N = 2\int_0^{L_e} n(x) dx$ and $n(L_e) = 0$ gives
\begin{eqnarray}
L_e & = & 1.15d^{2/5}D^{3/5}N ^{3/5},
\nonumber \\
\mu_0 & = & 1.32 e^2d^{4/5}D^{-9/5}N^{6/5},
\nonumber
\end{eqnarray}
so that, as announced earlier, the condition $N/ L_e\ll d^{-1}$ is equivalent to $N\ll\left(D/d\right)^{3/2}$.  The corresponding capacitance $C = e^2(d\mu_0/dN)^{-1}$ is given by
\begin{equation}
C \approx 0.629d\left(\frac{D}{d}\right)^{3/2}\left[\frac{N}{(D/d)^{3/2}}\right]^{-1/5}
\label{eq:ClowD}
\end{equation}

Thus, in the low density regime, the system exhibits an increase in the capacitance at small electron density, as in the case of a uniform electron system, due to the high electronic compressibility associated with the strong screening of the electron-electron interaction.  However, the growth of the capacitance is largely muted by the shrinking effective length, and retains only a very mild dependence, $C \sim N^{-1/5}$.

\begin{figure}
\begin{center}
\includegraphics[width=0.5\textwidth]{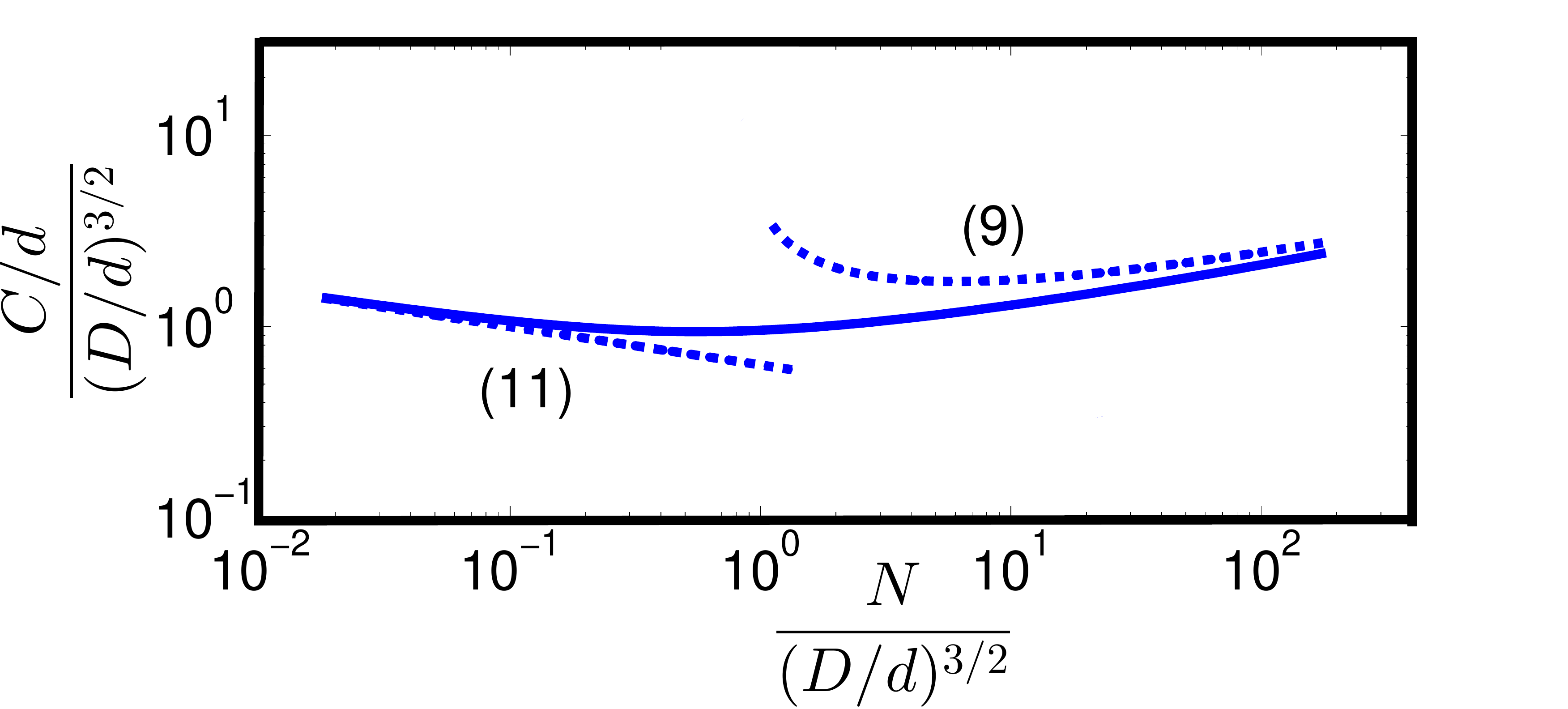}\\
\caption{Graph of capacitance $C$ as a function of electron number $N$ in the presence of a confining potential. $C$ is expressed in units of $d(D/d)^{3/2}$ and $N$ is normalized by $(D/d)^{3/2}$. The solid line shows the numerical result for capacitance. The dashed lines represent the asymptotic behaviors described by Eqs.\ (\ref{eq:ChighD}) and (\ref{eq:ClowD}).}
\label{fig:6}
\end{center}
\end{figure}

A plot of $C(N)$ is presented in Fig.\ \ref{fig:6} showing the different regimes of behavior. Notably, the capacitance no longer varies monotonically with $N$, first decreasing with $N$ in the low density limit and then increasing again at high density.

\begin{figure}
\begin{center}
\includegraphics[width=0.5\textwidth]{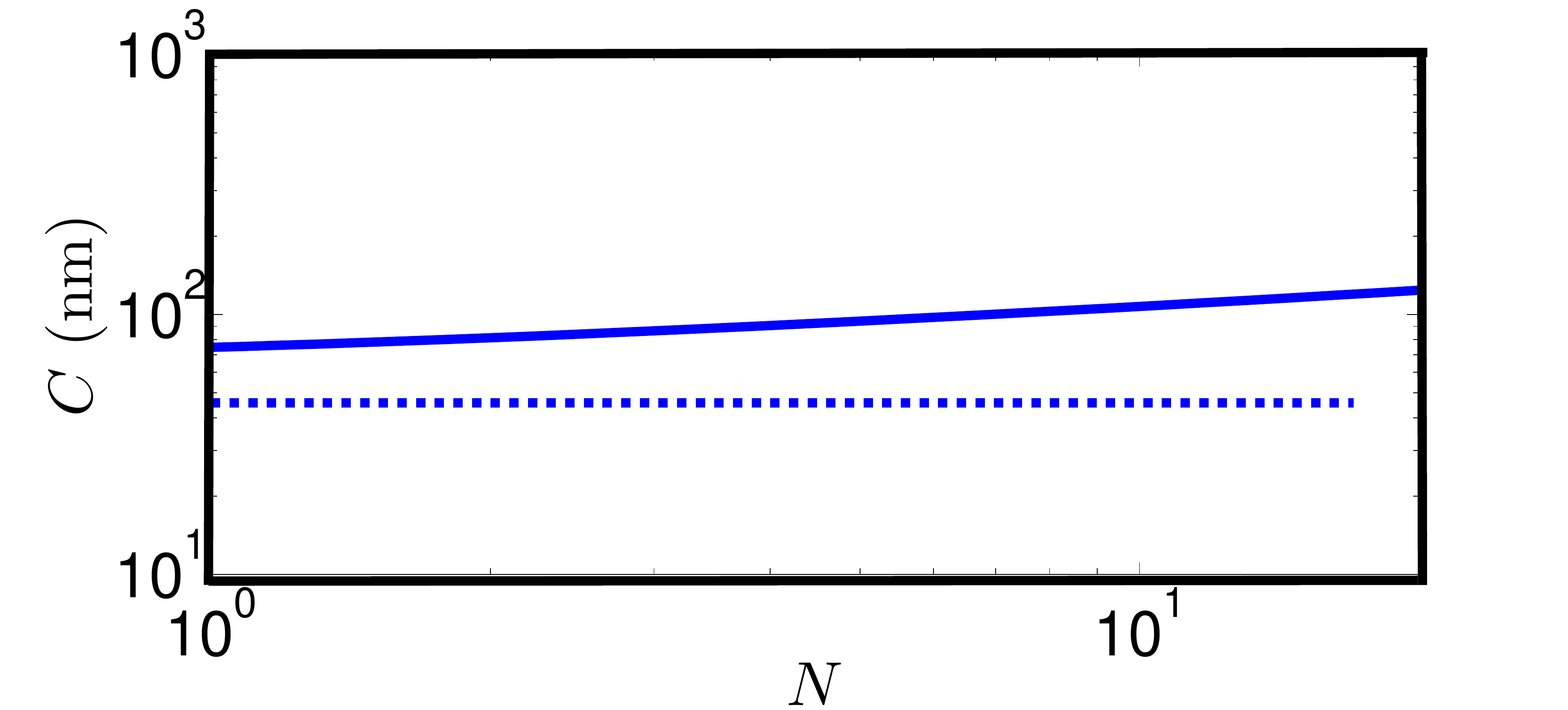}\\
\caption{A comparison between the predicted behavior of $C(N)$ and the experiments of Ref. \onlinecite{waissman_realization_2013}. The solid line is the prediction of the theory derived in Sec.~\ref{sec:withconf}. The dashed line represents the experimentally-measured capacitance $C_m$.}
\label{fig:7}
\end{center}
\end{figure}

This theoretical result is compared to the experimental measurements of Ref.\ \onlinecite{waissman_realization_2013} in Fig.\ \ref{fig:7}.
The results are qualitatively consistent with each other, suggesting that the confining potential provides an adequate explanation of the small, relatively constant capacitance seen in experiment.  A more accurate comparison to experiment may require more careful modeling of the electrostatic potential created by the experimental setup.

\section{Corrections to Capacitance Due to Quantum Effects}
\label{sec:variational}

So far, our calculations of the capacitance have employed a purely classical approximation, in which the quantum kinetic energy of electrons is ignored.  This approximation is generally justified in the strongly-interacting regime, where quantum effects provide only a small correction to the total energy.  Nonetheless, in this section we carefully justify our previous results by performing a calculation in which quantum mechanical effects are properly taken into account.

Our goal is to accurately estimate the ground state energy $E$ of the interacting electron system as a function of the total electron number $N$; the capacitance can then be found by Eq.\ (\ref{eq:Cdef}).  In other words, we seek an accurate estimate of the lowest-energy eigenvalue of the Hamiltonian
\be
H = -\sum_i\frac{\hbar^2\nabla_i^2}{2m} + \frac{1}{2}\sum_{i\neq j}V(r_{ij}).\\
\nonumber
\ee
(The additional interaction between an electron and its own image charge does not enter the capacitance, and can be ignored.) In order arrive at an estimate of $E$, we use a variational method, in which the expectation value of the Hamiltonian, $\langle H \rangle_\lambda = \langle \Phi_\lambda | H | \Phi_\lambda \rangle$, is calculated for a wide set of variational wavefunctions $\{ |\Phi_\lambda \rangle \}$, where $\lambda$ denotes some variational parameter that labels the different wavefunctions in the set.  The ground state energy $E$ is associated with the minimum value of $\langle H \rangle_\lambda$ across all possible values of $\lambda$.

Following Ref.\ \onlinecite{fogler_2005}, we use for our variational wavefunctions the eigenstates of the exactly-solvable Calogero-Sutherland model (CSM) \cite{calogero_1969, *sutherland_1971}, so that the variational parameter $\lambda$ describes the CSM interaction constant; $\lambda = 1$ corresponds to the non-interacting limit, while $\lambda \rightarrow \infty$ corresponds to a perfectly-ordered crystal.  In the limit of large system size, the CSM eigenstates have a kinetic energy per particle given by\cite{romer_2001}
\begin{displaymath}
\epsilon_k(n,\lambda)=\frac{\pi^2\hbar^2n^2}{6m}\frac{\lambda^2}{2\lambda-1},
\end{displaymath}
while the interaction energy per particle is
\begin{displaymath}
\epsilon_{int}(n,\lambda)=\frac{n}{2}\int V(r)g(\lambda,r)dr.
\end{displaymath}
Here, $g(\lambda,r)$ is the pair distribution function associated with the eigenstate $\Phi_\lambda$.  While $g(\lambda, r)$ has been studied in some detail analytically \cite{shashi_2012, sorokin_2006}, closed-form expressions are known only for the special values $\lambda=1/2,\,1,\,2,$ and $\infty$.\cite{romer_2001}  Thus, following Ref.\ \onlinecite{fogler_2005}, we evaluate $\epsilon_{int}(n, \lambda)$ at these special values, and then interpolate to other values of $\lambda$ by making a third-order polynomial fit as a function of $1/\lambda$.

The ground state energy per electron, $\epsilon(n)$, for a given density $n$ is equated with the minimum of the function $\epsilon_k(n, \lambda) + \epsilon_{int}(n, \lambda)$ across all values of $\lambda$.  The chemical potential is then calculated as $\mu = d(n \epsilon(n))/dn$.  For the case without a confining potential, the capacitance is equated with $e^2 L (dn/d\mu)$, as implied by Eq.\ (\ref{eq:Cdef}).  For the case with a confining potential, the capacitance is found by calculating the electrochemical potential associated with a given total electron number, $\mu_0 = \mu(n(x)) + e^2 D^2/x^3$, as explained in Sec.\ \ref{sec:withconf}, and then setting $C = e^2 (dN/d\mu_0)$.

\begin{figure}[h]
\includegraphics[width=0.5\textwidth]{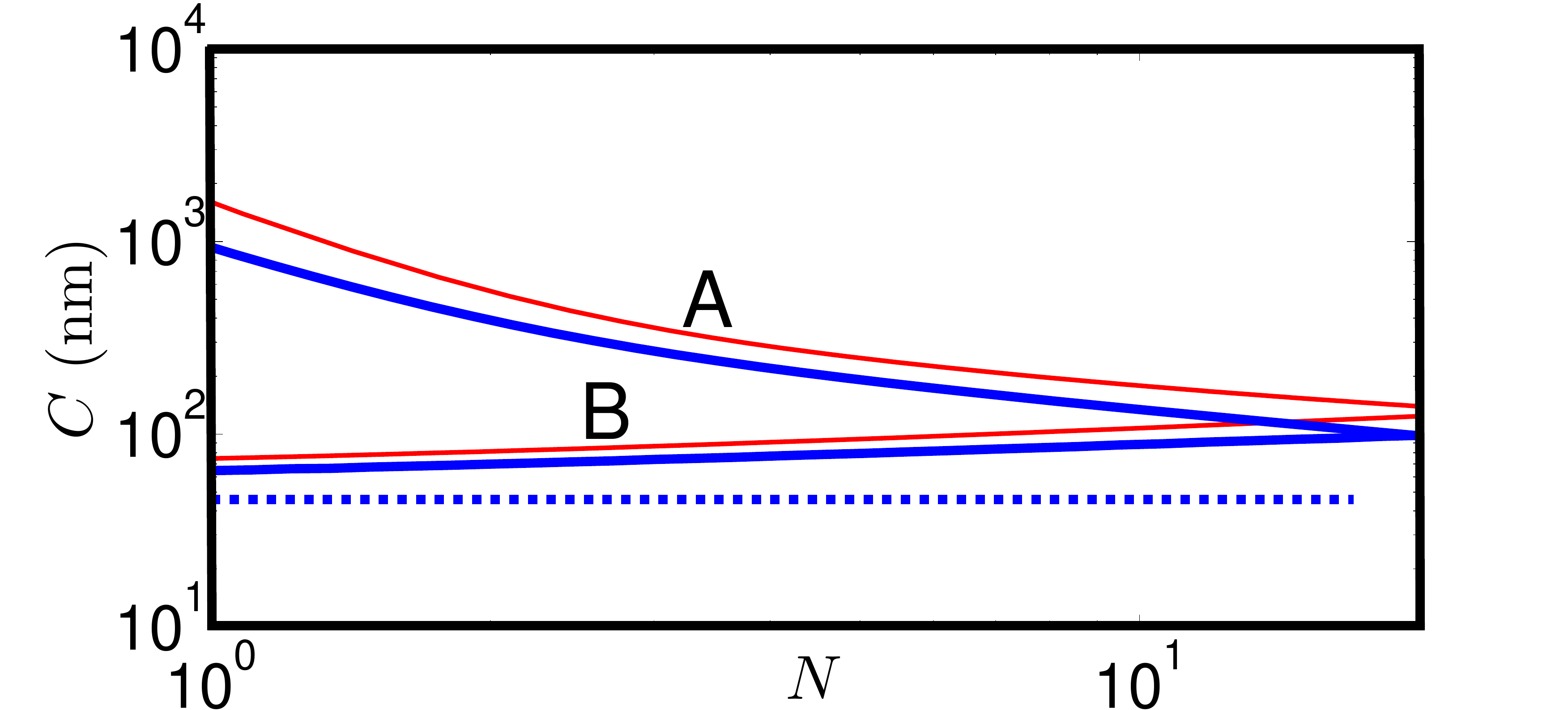}\\
\caption{(Color online) Plot of the capacitance $C$ against the number of electrons $N$. The two solid lines of group A are results for the case without a confining potential. The thick line (blue) is the result using the variational method presented in Sec.\ \ref{sec:variational}, while the thin line (red) is calculated using the classical approach in Sec.\ \ref{sec:noconf}. The two solid lines of group B have the same meaning but are for the case with a confining potential. The dashed line (blue) indicates the experimentally-measured value.}\label{fig:10}
\end{figure}

\begin{figure}[h]
\begin{center}$
\begin{array}{c}
\includegraphics[width=0.5\textwidth]{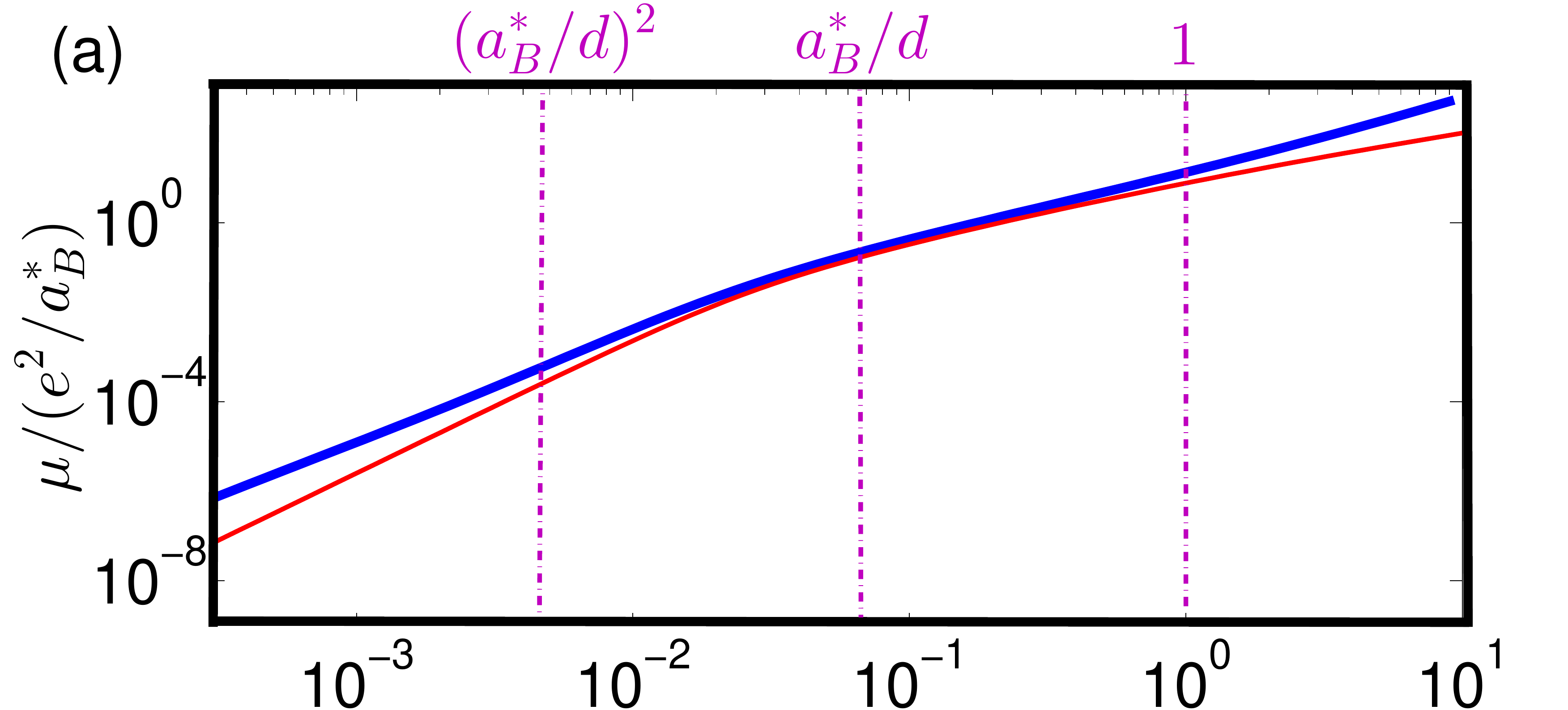}\\
\includegraphics[width=0.5\textwidth]{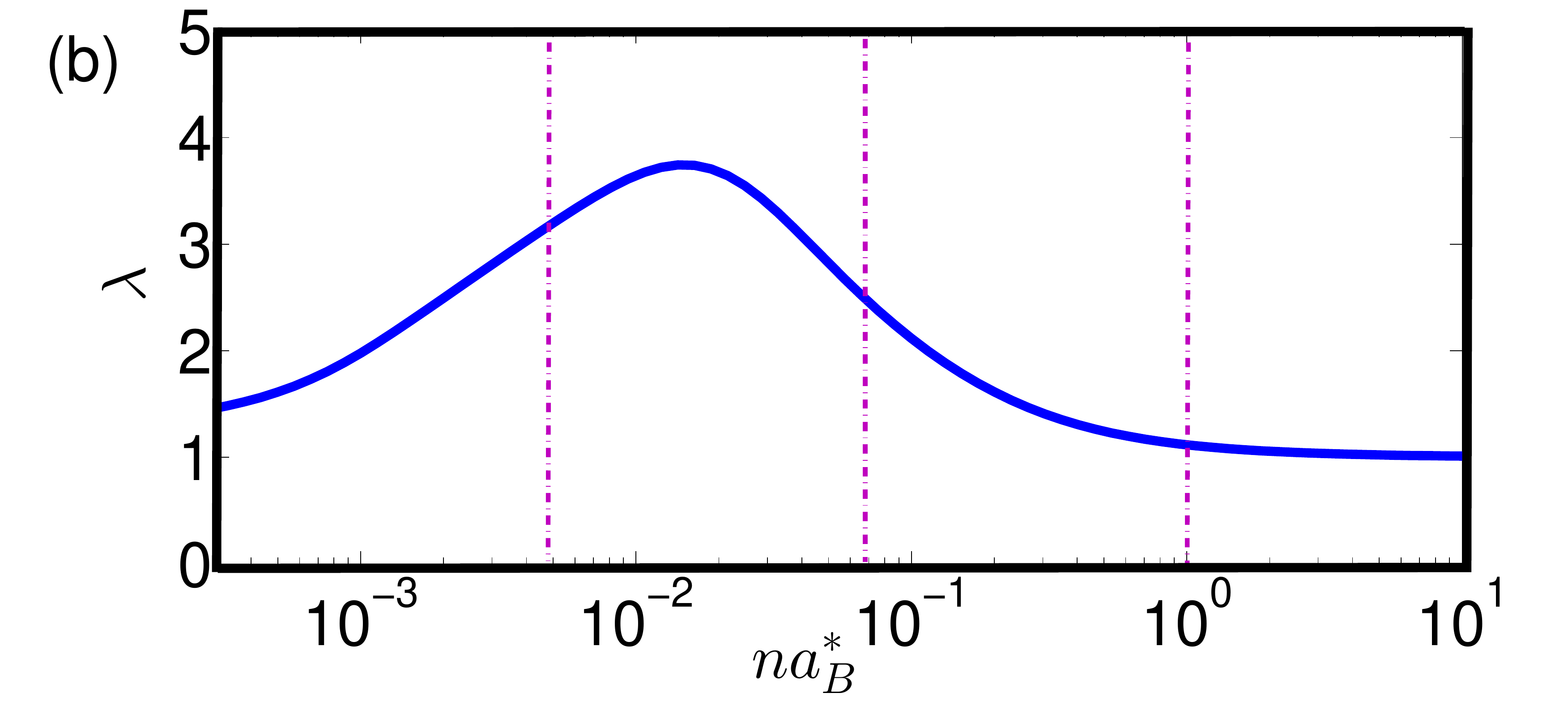}\\
\end{array}$
\caption{(Color online) Variational results for the chemical potential $\mu$ and the variational parameter $\lambda$ as a function of the dimensionless concentration $na_B^*$ at $d=15a_B^*$. (a) The thick solid line (blue) represents $\mu(n)$ in units of $e^2/a_B^*$. The thin solid line (red) corresponds to the result obtained by the classical model in Sec.\ \ref{sec:noconf}. (b) The variational parameter $\lambda$ associated with the minimum energy wavefunction at a given density $n$. Vertical dotted lines denote different regimes of electron density; from left to right: the low-density weakly interacting regime at $n\ll a_B^*/d^2$, the dipole-interacting Wigner-crystal-like regime at $a_B^*/d^2\ll n\ll 1/d$, the unscreened Wigner-crystal-like regime at $1/d\ll n\ll 1/a_B^*$, and the high-density weakly interacting regime at $1/a_B^*\ll n$. In the crystal-like regimes, the difference between variational and classical results for $\mu$ is small and $\lambda$ is relatively big, denoting strong correlations. In the weakly-interacting regimes, the difference in $\mu$ grows and $\lambda$ approaches unity.}\label{fig:9}
\end{center}
\end{figure}

The resulting capacitance is plotted in Fig.\ \ref{fig:10} for the range of experimental parameters explored in Ref.\ \onlinecite{waissman_realization_2013}.  As can be seen, the results closely follow those derived from the simple classical description of Sec.\ \ref{sec:noconf} and \ref{sec:withconf}, suggesting that quantum effects indeed provide only a small correction within the experimental range of electron density.  One can also see in Fig.\ \ref{fig:9} the non-monotonic dependence of the correlation strength on the electron density.  Indeed, as the electron density is first reduced beginning at large density, the electrons become more strongly correlated as the typical ratio $\epsilon_{int}/\epsilon_k \sim (e^2n)/(\hbar^2 n^2/m) \sim (na_B^*)^{-1}$ is increased, and correspondingly the value of $\lambda$ associated with the variational wavefunction increases.  On the other hand, at small enough density that $n \ll a_B^*/d^2$, the interactions between electrons is dipolar, so that the ratio $\epsilon_{int}/\epsilon_k \sim (e^2 d^2 n^3)/(\hbar^2 n^2/m) \sim n d^2 /a_B^*$.  Consequently, the correlations between electrons become weak again as the density is reduced further, and the value of $\lambda$ falls.

\section{Conclusion}
\label{sec:conclusion}

In this paper we have studied the 1D correlated electron gas in a gated carbon nanotube device. We calculated the capacitance of the system in different density regimes, which exhibit different electronic behaviors. For a spatially uniform system, the electrons occupy a weakly-interacting phase at high electron density, and correspondingly the capacitance is constant and relatively small.  As the electron density is lowered, the electrons acquire strong, Wigner crystal-like correlations, and the capacitance becomes weakly enhanced.  At sufficiently low electron density that $nd \ll 1$, the electron-electron interactions are strongly screened by the metal gate, and the capacitance increases sharply.

\begin{figure}[h]
\begin{center}$
\begin{array}{c}
\includegraphics[width=7cm, height=3.8cm]{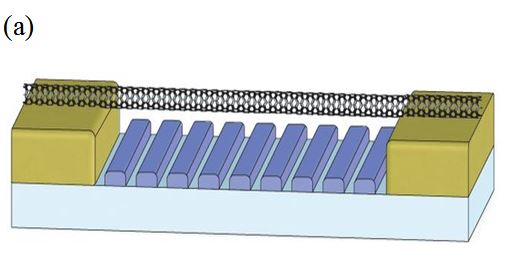}\\
\includegraphics[width=7cm,height=4cm]{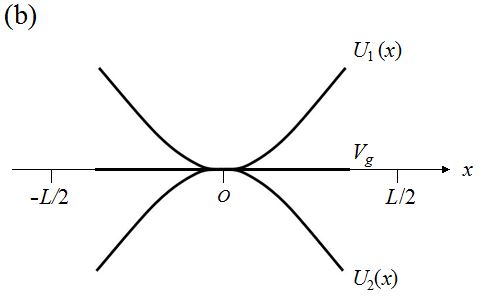}
\end{array}$
\caption{(Color online) Possible realization of constant potential along the CNT by using multiple split gates set at different gate voltages.  (a) A schematic of the device setup, taken (with permission) from Ref. \onlinecite{waissman_realization_2013}. (b) By adjusting the voltages on the multiple gates, one creates a potential $U_2(x)$ which compensates the confining potential $U_1(x)$ caused by the source/drain-gate fields, resulting in a uniform potential $V_g=U_1(x)+U_2(x)$ across most of the CNT.}
\label{fig:11}
\end{center}
\end{figure}

These predictions for the uniform system, however, are qualitatively different from existing experimental measurements, which show a capacitance that is roughly independent of electron concentration and much smaller in magnitude.  This discrepancy can be understood by accounting for the effects of a confining electrostatic potential that tends to push electrons toward the center of the CNT, and thereby reduce the effective length of the capacitor at small electron density. This shrinkage of the effective capacitor length greatly reduces the capacitance at small electron density, and results in an estimate for the capacitance that is much more consistent with experiment.

If, in the future, the confining potential could be eliminated, the dramatic increase in capacitance at low density would be restored, and this would permit more sensitive probing of correlated ``dipolar electron" physics. Further studies should thus consider how to realize this reduced confining potential in the experimental environment. One possible way is to use multiple gates with different voltages on each one, as depicted schematically in Fig.\ \ref{fig:11}, in order to tune the electrostatic potential along the CNT. Such devices have already been fabricated \cite{waissman_realization_2013}, but careful attempts to realize a constant potential have not yet been reported. Such a constant potential will be achieved when a parabolic potential \emph{maximum} simulated by multiple gates precisely balances out the parabolic minimum created by the source/drain-gate potentials.

Finally, as noted above, we have so far not considered the effect of finite temperature, while the experiment of Ref.\ \onlinecite{waissman_realization_2013} was performed at $T=4$ K. Nonetheless, throughout the experimental range of electron density, the typical electron-electron interaction energy  $\varepsilon$ is such that $\varepsilon/k_B\gtrsim 96$ K$\, \gg4$ K, and so our zero temperature approximation is justified.

\acknowledgments

We are grateful to S. Ilani, J. Waissman, M. M. Fogler and K. A. Matveev for helpful discussions. This work was supported partially by the National Science Foundation through
the University of Minnesota MRSEC under Award Number DMR-1420013. Work at Argonne National Laboratory is supported by the U.S. Department of Energy, Office of Science under contract no. DE-AC02-06CH11357.

\appendix

\section{Potential Produced by Three Coplanar Metallic Strips} \label{App:AppendixA}

\begin{figure}[h]
\begin{center}
\includegraphics[width=0.5\textwidth]{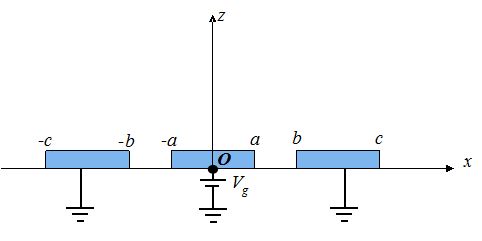}\\
\caption{A schematic graph of three coplanar metallic strips separated by a distance $b-a$ and held at different potentials.}
\label{fig:8}
\end{center}
\end{figure}

To calculate the potential $V$ produced by three coplanar metallic strips placed next to each other, we first consider them as separated from each other by a certain distance $b-a$ and then take the limit of $b-a=0$ to get the final result. The middle strip is set at a potential $V_g$ while the side ones are at $V=0$. The electric field is independent of the longitudinal coordinate (perpendicular to the CNT and parallel to the gate electrode), so that solving for the potential can be reduced to a two-dimensional electrostatic problem, as shown in Fig.\ \ref{fig:8}. Therefore along the $x$ axis, we have a field
\be
E_x(x)=\int^\infty_{-\infty}\frac{2\rho(x')dx'}{x-x'}
\nonumber
\ee
where $\rho(x')$ is the surface charge density on the metallic strips at $x=x'$.

Then using Kramers-Kronig relations, we can define a complex function $f(\zeta)$, $\zeta=x+iz$, which satisfies Re$f(x)=E_x(x)$ and Im$f(x)=-2\pi\rho(x)$. Since
\be
\begin{array}{l}
\,\,\,\,\rho(x)=0,\quad\quad a<|x|<b\textrm{ or }|x|>c\\
E_x(x)=0,\quad\quad b<|x|<c\textrm{ or }|x|<a
\end{array}
\nonumber
\ee
we guess $f(\zeta)=\frac{A}{\sqrt{(x^2-a^2)(x^2-b^2)(x^2-c^2)}}$. Then,
\be
V_g=\int^b_a\frac{A}{\sqrt{(x^2-a^2)(b^2-x^2)(c^2-x^2)}}dx.
\nonumber
\ee

Taking $b=a(1+\xi)$, and expanding for $\xi\ll1$, we have
\be
\begin{array}{ll}
V_g&\approx\frac{A}{a\sqrt{c^2-a^2}}\frac{\pi}{8}\\
&\\
A&=\frac{8a\sqrt{c^2-a^2}V_g}{\pi}\\
\end{array}
\nonumber
\ee

Using the uniqueness theorem, we know this is the only solution to this configuration. Then on the middle strip, the charge density is
\be
\begin{array}{ll}
\rho(x)&\approx\frac{4a\sqrt{c^2-a^2}V_g}{\pi^2(a^2-x^2)\sqrt{c^2-x^2}},\quad |x|<a\\
\end{array}
\nonumber
\ee
and the electric field in the vertical direction, $E_z(x,0^+)=2\pi\rho(x)$, is
\be
E_z(x,0^+)\approx\frac{8\sqrt{c^2-a^2}V_g}{\pi ac}+\frac{8\sqrt{c^2-a^2}V_g}{\pi ac}\left(\frac{1}{a^2}+\frac{1}{2c^2}\right)x^2.
\nonumber
\ee
where $x\ll a$.
So near the surface of the middle metallic strip at a height of $d$, the electric potential is
\be
V(x,d)\approx V_g-E_z(x,0^+)d=V(0)-\frac{e}{D^3}x^2
\nonumber
\ee
where
\be
\begin{array}{rl}
V(0)&=V_g-\frac{8d\sqrt{c^2-a^2}V_g}{\pi ac},\\
D&=\left[\frac{\pi ac e^2}{8d\sqrt{c^2-a^2}eV_g}\frac{1}{1/a^2+1/(2c^2)}\right]^{\frac{1}{3}}.
\end{array}
\nonumber
\ee

If we denote the width of the middle strip as $L$ and that of the side ones as $s$, then $a=L/2,\,c=L/2+s$, and
\be
\begin{array}{rl}
V(0)&=V_g-\frac{32d\sqrt{s(L+s)}V_g}{\pi L(L+2s)},\\
D&=\left[\frac{\pi (L/s+2) e^2}{64d\sqrt{L/s+1}eV_g}\frac{1}{2+1/(1+2s/L)^2}\right]^{\frac{1}{3}}L.\\
\end{array}
\nonumber
\ee

Inserting the experimental parameters from Ref.\ \onlinecite{waissman_realization_2013} ($d = 130$\,nm, $L = 880$\,nm, $Vg \approx 0.5-1$\,V, and $s \approx 300$\,nm) gives $D \approx 72-90$\,nm.

\bibliography{CNT}

\end{document}